\documentclass[11pt,twoside]{article}
\usepackage{asp2014}

\aspSuppressVolSlug
\resetcounters

\begin{document}

\title{Diffusion coefficients in white dwarfs}
\author{D. Saumon,$^1$ C.E. Starrett,$^1$ and J. Daligault$^1$}
\affil{$^1$Los Alamos National Laboratory, Los Alamos, NM, USA; \email{dsaumon@lanl.gov}}

\paperauthor{D. Saumon}{dsaumon@lanl.gov}{}{Los Alamos National Laboratory}{X-Computational Physics Division}{Los Alamos}{NM}{87545}{USA}
\paperauthor{C. E. Starrett}{starrett@lanl.gov}{}{Los Alamos National Laboratory}{X-Computational Physics Division}{Los Alamos}{NM}{87545}{USA}
\paperauthor{J. Daligault}{daligaul@lanl.gov}{}{Los Alamos National Laboratory}{Theoretical Division}{Los Alamos}{NM}{87545}{USA}

\begin{abstract}
Models of diffusion in white dwarfs universally rely on the coefficients
calculated by \citet{paquette86}.  We present new
calculations of diffusion coefficients based on an advanced 
microscopic theory of dense plasmas and a numerical simulation approach
that intrinsically accounts for multiple collisions.  Our method is validated against a state-of-the-art 
method and we present results for the diffusion of carbon ions in a helium plasma.
\end{abstract}

\section{Introduction}

The spectral type, atmospheric composition, and spectral evolution of white dwarfs (WDs) is largely determined 
by the diffusion of elements in the envelope of the star. 
Modern models of WDs universally rely on 
the diffusion coefficients of \citet{paquette86}. Advances in the theory of dense plasmas and 
in computational capabilities in the last three decades motivate a re-evaluation of the coefficient of ionic 
diffusion in WD envelopes, particularly under physical conditions that are challenging to model.
Our physical model and computational approach are very different from those of \citet{paquette86}.
We have developed a new microscopic model of dense, partially ionized, multi-component 
plasmas. When combined with a molecular dynamics simulation of the ions in the plasma mixture, 
the diffusion coefficient can be evaluated.  We discuss our model and methods and contrast them
with the approach of \citet{paquette86}. Finally, we present our first calculation of the diffusion
coefficient in WDs for a mixture of carbon and helium.

\subsection{Microscopic model for dense plasmas}

\citet{ss13} and \citet{ss_hedp} describe a detailed and realistic microscopic model for dense, partially ionized, multi-component 
plasmas that accounts for strong interactions between ions, partial electron degeneracy, bound states, 
pressure ionization, and non-linear electron screening.  For clarity of the following discussion, we will consider a plasma consisting
of a single species of ions and free electrons although the model can describe a mixture of
species equally well \citep{ssdh_mix}.  The model consist of two main elements which are coupled.  First the ions are modeled 
within the framework of an average atom model. This assumes that all ions in the plasma are identical
and imposes spherical symmetry on the distribution
of bound and free electrons around the nucleus.  The electronic states (bound and free) are obtained
by solving the Schr\"odinger (or Thomas-Fermi) equation within the density functional formalism.  The potential in which the
electrons evolve accounts for the central nucleus, the other electrons belonging to the ion, and
the distribution of other charges in the surrounding plasma. Thus 
non-ideal effects on the electronic structure of the ion (such as pressure ionization) are naturally included.
This part of the model provides the average ion charge and the distribution of the continuum electrons around the
ion (the screening cloud of electrons).  
The second part of the model uses those two quantities to calculate the structure of the surrounding plasma. 
In the dense plasmas found in WDs, the ion-ion interactions can be very strong. While the fluid is disordered, the
positions of the particles are not random but correlated.  
The statistical average of the relative positions of particles in the plasma is described by correlation functions
that are calculated with the integral equations formalism of fluid theory.  This
in turn provides the external plasma distribution needed to solve the Schr\"odinger equation of the average atom.
A consistent solution of the two coupled parts of the model is obtained by numerical iteration and provides the ion-ion
interaction potential to be used in molecular dynamics simulations (see below). This model is physically
self-consistent and has no adjustable parameters.

\subsection{Evaluation of the diffusion coefficient}

All the equations describing the above plasma model are time independent. On the other hand, diffusion is
an inherently dynamic process that describes how a particle drifts from an initial position.  For the purpose of evaluating 
the coefficient of ionic diffusion, we resort to the numerical simulation method of molecular dynamics (MD, \citet*{md_book}).
In a MD simulation, a fluid of classical particles interacting with a pair potential at a given density and temperature is 
represented by $N$ particles (typically a few thousands) in a cubic box of fixed volume. After the position and velocity of each 
particle in the box have been initialized, the total force on each particle is calculated from the sum of the pair interactions. 
The position and velocity of each particle are updated by integrating Newton's law $F=ma$ over a small time step. 
The forces are then recalculated. This simple scheme is repeated to let the system evolve in time.
The diffusion coefficient can be evaluated from the trajectories of the
particles in the simulation.  An infinite system is approximated by using periodic boundary conditions at the 
surface of the box. 

At the microscopic level, diffusion describes the drift in the position of a particle from 
its initial position due to its thermal motion and collisions with other particles. From the 
particle's trajectory $\vec r(t)$, the coefficient of self-diffusion is expressed as \citep{hmcd}
\begin{equation}
  D=\lim_{t\to\infty} \frac{1}{6t} \big< \big| \vec r(t) - \vec r(0) \big|^2 \big>
  \label{self-diff}
\end{equation}
where $\vec r(0)$ is its initial position 
and $<>$ represent the thermal average over all particles in the simulation. Equation (\ref{self-diff}) says
that after a large number of collisions, the particle has traveled a distance that is proportional to
$\sqrt t$ (the well-known random walk result), and that the proportionality constant is six times
the diffusion coefficient. For a plasma of two species, which is of interest in WDs,
the coefficient of inter-diffusion can be obtained from $ D_{12}=J(x_2D_1 + x_1D_2) $,
where $D_i$ is the self-diffusion coefficient of species $i$, $x_i$ is its number fraction
($x_2=1-x_1$), and $J$ is the so-called thermodynamic factor \citep{hjmcd}. For ideal gases and for
trace diffusion ($x_1\to 0$), $J=1$. Otherwise, our calculations show that $J$ can be as large as 1.5. 
This expression for $D_{12}$ is an approximation that neglects cross-correlation terms. It
has been shown to be fairly accurate to a few percent for some systems \citep{hjmcd,daligault12} but should to be 
re-evaluated for the conditions of interest here \citep{hax14}.

\subsection{Comparison with the approach of Paquette et al. (1986)}

Our approach to evaluate the coefficient of inter-diffusion in WDs is compared with that of
\citet{paquette86} in Table \ref{table1}.  For the diffusion
of a trace element (Ca, for example) in a plasma dominated by one species (typically H or He), the
most significant advantages of our calculation are 1) the ion-ion potential is calculated from a microscopic model of
the plasma and does not have an assumed functional form, 2) the use of
MD to evaluate the diffusion coefficient, which is not limited to low densities, and 3) the
self-consistency of the approach that uses a single model to generate the EOS and the diffusion coefficients.
Furthermore, for non-dilute mixtures our approach properly accounts for the thermodynamic factor $J$.

\begin{table}[!ht]
\caption{Comparison with the approach of Paquette et al. (1986).}
\smallskip
\begin{center}
{\small
\begin{tabular}{lcc}  
\tableline
\noalign{\smallskip}
Assumption / Approximation & Paquette et al. (1986) & This work \\
\noalign{\smallskip}
\tableline
\noalign{\smallskip}
Equation of state (EOS) &    Ideal gas               &  Fully non-ideal \\
\noalign{\smallskip}
Order of the collisions &  2-body collisions only    &   N-body collisions \\
\noalign{\smallskip}
Ion-ion pair potential  &  Static screened Coulomb   &  self-consistent with \\
        ~               &   ``Debye-H\"uckel''       &  plasma model \\
\noalign{\smallskip}
Diffusion coefficient   &  Collision integrals       &  Molecular dynamics \\
         ~              &  (theory for dilute gases) &   (valid for dense fluids) \\
\noalign{\smallskip}
Thermodynamic factor ($J$)    &  $J=1$               & From non-ideal EOS \\
\noalign{\smallskip}
\tableline\
\end{tabular}
}
\end{center}
\label{table1}
\end{table}

\subsection{Accuracy and application to carbon/helium plasmas in white dwarfs}

The diffusion of C in He at the core/envelope interface and its dredge up by the He 
convection zone is responsible for the presence of C in the spectra of DQ WDs \citep{pelletier86}. A change 
in the diffusion coefficient would affect the quantitative modeling of this process and estimates of 
the mass of the He layer in DQ stars.  

Thomas-Fermi Molecular Dynamics (TFMD, \citet*{zerah92}) is a state-of-the-art method for
simulating hot, dense plasmas. It explicitly accounts for the electrons of the plasma (with the 
Thomas-Fermi model) and the nuclei with a minimal set of approximations. TFMD is more accurate than 
the present model but is computationally too expensive to generate large tables of diffusion coefficients.
A comparison with TFMD validates our calculations of $D_{12}$
for C/He mixtures over a wide range of conditions to better than 5\% (Fig. \ref{Diff_in_WD}). 
Along a WD envelope profile, our diffusion coefficients for C/He are in excellent
agreement with those of \citet{paquette86} at depth ($\log T > 6.4$ or $\log q > - 6.9$) but are up to factors of 3 larger 
in the upper envelope (Fig. \ref{Diff_in_WD}).

\articlefiguretwo{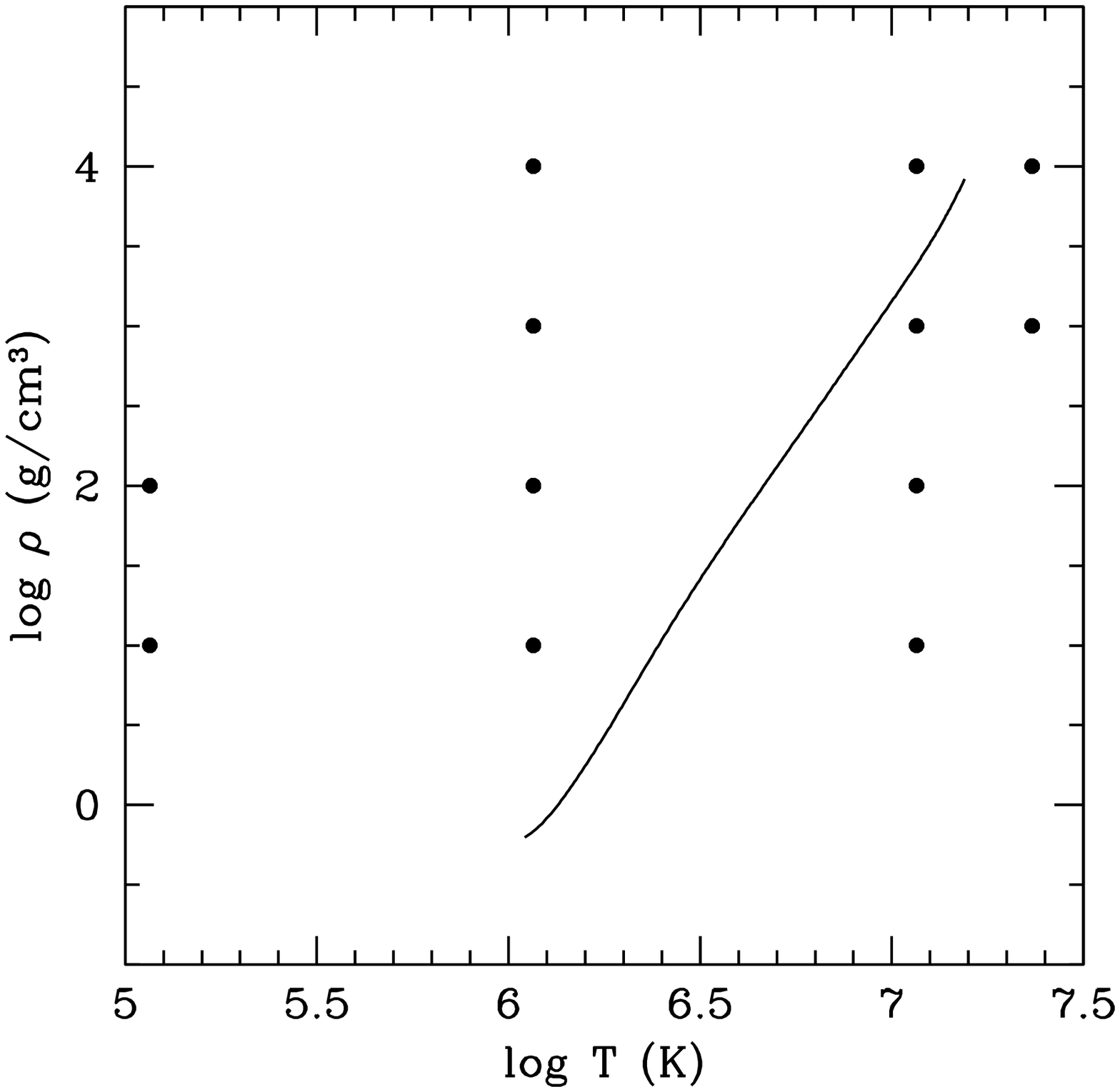}{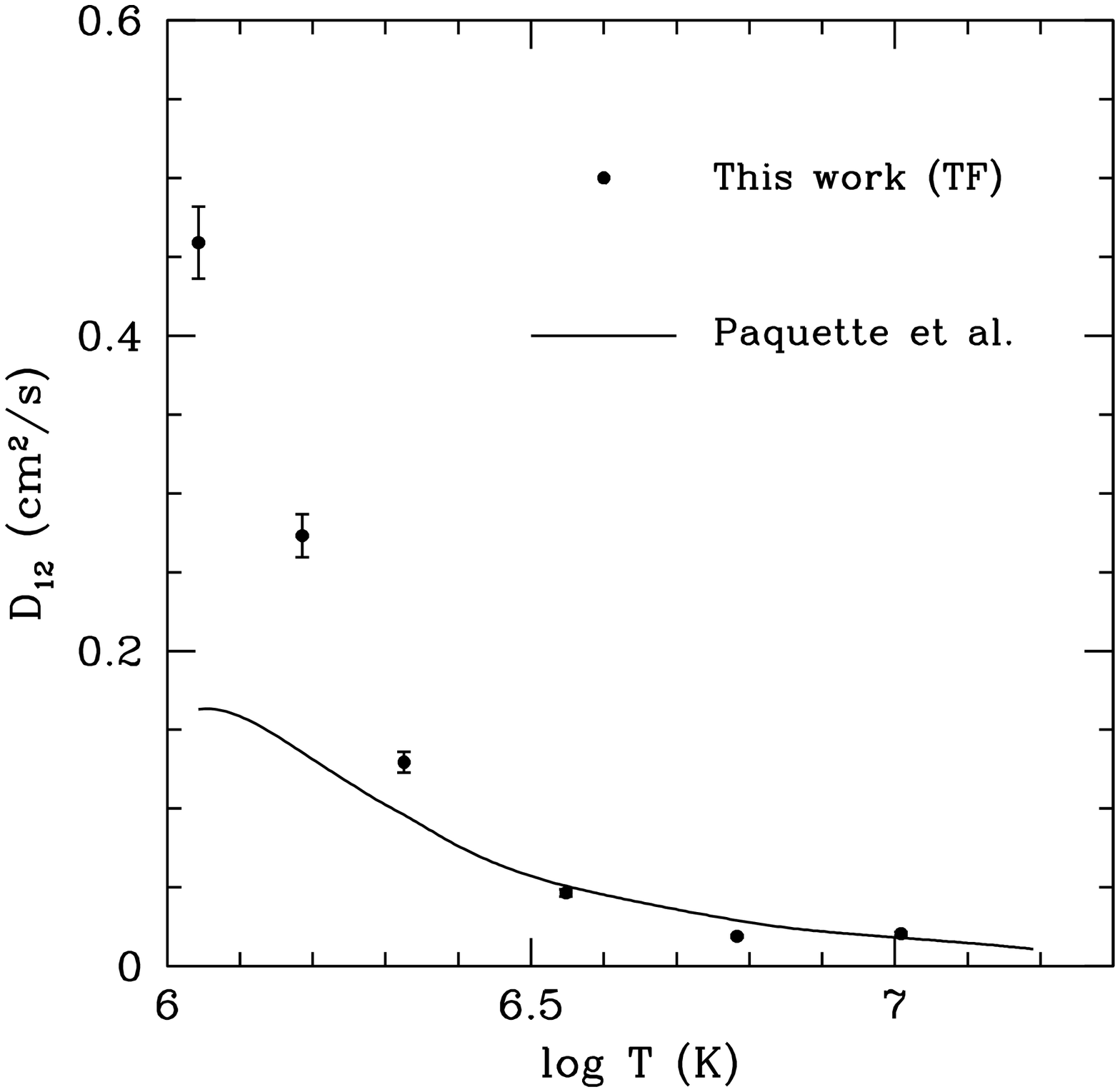}{Diff_in_WD}{\emph{Left:} Solid curve: Envelope profile of a
     DB WD model ($T_{\rm eff}=20000\,$K, $\log g=8$). Dots show the points where we validated our 
     approach against TFMD simulations. \emph{Right:} Comparison of the
     diffusion coefficient of a trace of C in He with \citet{paquette86} along the profile shown in 
     the left panel. Error bars are estimates of the uncertainty in our calculation ($\pm 5$\%).}

\subsection{Conclusions}

We have developed a realistic model of the plasmas found in the envelopes of WDs that provides a 
self-consistent description of the equation of state and diffusion coefficients. Our approach relaxes several 
of the approximations  of \citet{paquette86} that are not well justified in WD envelopes.  Our initial
results for inter-diffusion in C/He mixtures show that our calculations of the diffusion coefficient are accurate to better than 5\%.
Along a model envelope of a WD, we find excellent agreement with \citet{paquette86} at depth but
large deviations in the upper envelope of up to a factor of three.

\acknowledgements We thank Gilles Fontaine for providing the white dwarf envelope model shown in Figure \ref{Diff_in_WD} and
for valuable discussions. This work was performed under the auspices of the United States Department of Energy under 
contract DE-AC52-06NA25396 and was supported in part by the LANL LDRD program.


\begin{thebibliography}{}
\bibitem[\protect\astroncite{Allen \& Tidesley}{1989}]{md_book} Allen, M.P. \& Tidesley, D.J. 1989, Computer simulations of liquids, 2$^{\rm nd}$ Ed., 
                             (Oxford University Press, Oxford)
\bibitem[\protect\astroncite{Daligault}{2012}]{daligault12} Daligault, J. 2012, Phys. Rev. Lett., 108, 225004
\bibitem[\protect\astroncite{Hansen \& McDonald}{1986}]{hmcd} Hansen, J.-P. \& McDonald, I.R. Theory of simple liquids, 2$^{\rm nd}$ Ed., 
                             (Academic Press, London)
\bibitem[\protect\astroncite{Hansen, Joly \& McDonald}{1985}]{hjmcd} Hansen, J.-P., Joly, F. \& McDonald, I.R. 1985, Physica 132A, 472
\bibitem[\protect\astroncite{Haxhimali et al.}{2014}]{hax14} Haxhimali, T., Rudd, R.E., Cabot, W.H. \& Graziani, F.R. 2014, Phys. Rev. E, 90, 023104
\bibitem[\protect\astroncite{Paquette et al.} {1986}]{paquette86} Paquette, C., Pelletier, C., Fontaine, G. \& Michaud, G. 1986, \apjs, 61, 177
\bibitem[\protect\astroncite{Pelletier et al.}{1986}]{pelletier86} Pelletier, C., Fontaine, G., Wesemael, F. \& Michaud, G. 1986, \apj 307, 242
\bibitem[\protect\astroncite{Starrett \& Saumon}{2013}]{ss13} Starrett, C.E. \& Saumon, D. 2013, Phys. Rev. E, 87, 013104 
\bibitem[\protect\astroncite{Starrett \& Saumon}{2014}]{ss_hedp} Starrett, C.E. \& Saumon, D. 2014, High Ener. Dens. Phys., 10, 35 
\bibitem[\protect\astroncite{Starrett et al.}{2014}]{ssdh_mix} Starrett, C.E., Saumon, D., Daligault, J. \& Hamel, S.  2014, Phys. Rev. E, in press
\bibitem[\protect\astroncite{Z\'erah, Cl\'erouin \& Pollock}{1992}]{zerah92} Z\'erah, G.,  Cl\'erouin, J. \& Pollock, E.L.  1992, Phys. Rev. Lett., 69, 446 
      
\end{thebibliography}


\end{document}